\documentclass[12pt,preprint]{aastex}
\def\t0{\theta_{\circ}}

\def\be{\begin{equation}}
\def\en{\end{equation}}

\def\kms{kms$^{-1}$}

\newcommand{\msun}{$M_{\odot}$}
\begin{document}

\shorttitle{Accretion in Young Brown Dwarfs}
\shortauthors{Jayawardhana et al.}

\title{Probing Disk Accretion in Young Brown Dwarfs}

\author{Ray Jayawardhana}
\affil{Department of Astronomy, University of Michigan, 830 Dennison, Ann Arbor, MI 48109}
%\email{rayjay@astro.berkeley.edu}

\author{Subhanjoy Mohanty}
\affil{Harvard-Smithsonian Center for Astrophysics, 60 Garden St., Cambridge, MA 02138}

\and

\author{Gibor Basri}
\affil{Department of Astronomy, University of California at Berkeley, Berkeley, CA 94720}

\begin{abstract}
We present high-resolution optical spectra of 15 objects near or below
the sub-stellar limit in the Upper Scorpius and $\rho$ Ophiuchus star-forming
regions. These spectra, obtained with the HIRES instrument on the Keck I 
telescope, are used to investigate disk accretion, rotation and activity in
young very low mass objects. We report the detection of a broad, asymmetric 
H$\alpha$ emission line in the $\rho$ Oph source GY 5 which is also known
to harbor mid-infrared excess, consistent with the presence of an accreting 
disk. The H$\alpha$ profiles of the Upper Sco objects suggest little or 
no on-going accretion. Our results imply that if most brown dwarfs are 
born with disks, their accretion rates decrease rapidly, at timescales 
comparable to or smaller than those for T Tauri disks. The Upper Sco brown
dwarfs appear to be rotating faster than their somewhat younger counterparts
in Taurus, consistent with spin-up due to contraction following disk 
unlocking. The H$\alpha$ activity 
is comparable to saturated activity levels in field M dwarfs with similar 
spectral type and rotation rates. Comparison of our data with published 
(albeit lower-resolution) spectra of a few of the same objects from other 
epochs suggests possible variability in accretion/activity indicators. 

\end{abstract}


\section{Introduction}

In their pre-main-sequence phase, low mass stars accrete high angular momentum 
material from a circumstellar disk.  The star remains a comparatively slow 
rotator, however, apparently due to disk-locking, i.e., magnetic braking by 
large-scale stellar field lines that thread the disk.  In the classical T
Tauri phase, the fields may also mediate accretion by truncating the inner 
disk and channeling in infalling material (K\"onigl 1991; Ostriker \& Shu 
1995).  While many independent observations support this picture for stars 
with masses $>$0.2\msun, whether it applies to objects at or below the 
hydrogen-burning limit is only now being explored. 

The formation mechanisms of brown dwarfs are still open to debate. Most 
recently, Padoan \& Nordlund (2002) have argued that brown dwarfs form in
the same way as more massive stars, via `turbulent fragmentation'. Reipurth 
\& Clarke (2001), on the other hand, suggested that brown dwarfs are 
stellar embryos ejected from newborn multiple systems before they accreted
sufficient mass to eventually start hydrogen burning. In this model, 
dynamical interactions are expected to prune their disks. Studies of disk 
accretion, rotation and activity in young brown dwarfs could help distinguish 
between these scenarios. In particular, it is of considerable interest 
to investigate whether some or all young sub-stellar objects undergo a T 
Tauri-like phase and if so how long that phase lasts. 

Accretion signatures have been seen already in high-resolution spectra of a 
few very low mass sources. Muzerolle et al. (2000) detected an asymmetric 
H$\alpha$ line profile in the M6 object V410 Anon 13 in Taurus and 
used magnetospheric accretion models to show that it is accreting from a 
disk but at a much lower rate than that found in higher mass stars. White
\& Basri (2002) found two more Taurus M5.5-M6.5 objects with H$\alpha$
profiles similar to other classical T Tauri stars.

Here we present high-resolution optical spectroscopy of a much larger sample 
of objects near or below the sub-stellar limit in the nearby ($\sim$ 150 pc) 
Upper Scorpius and $\rho$ Ophiuchus star-forming regions. Our targets are
sources with M5 or later spectral types from the surveys of Ardila, Mart\'in,
\& Basri  (2000) for Upper Sco and Wilking, Greene \& Meyer (1999) for 
$\rho$ Oph. Our goals are to search for broad, asymmetric 
H$\alpha$ lines indicative of disk accretion, to measure rotational 
properties as denoted by {\it v~sin~i}, and to investigate 
chromospheric activity, in these young very low mass objects in comparison
to their older counterparts in the field and to (higher-mass) T Tauri stars. 

In a subsequent paper (Mohanty et al., in prep), we will use these optical
spectra, together with observed photometry and theoretical isochrones, to 
derive effective temperatures, gravities, masses and ages for our sample 
objects. For the purposes of this paper, we assume an age of $\sim$ 1 Myr 
for $\rho$ Oph and $\sim$ 5 Myr for Upper Sco (Preibisch \& Zinnecker 1999). 
If we also assume that the effective temperatures of our young M5-M8 objects
are comparable to those of field M dwarfs of the same spectral type 
($\sim$ 3000-2500 K, similar to what is derived by Wilking, Greene \& Meyer 
1999 for their $\rho$ Oph sample), then theoretical models (Chabrier et al. 
2000) imply masses of $\sim$ 0.1 -- 0.02 \msun. Thus, our sample spans a mass  
range from the stellar/sub-stellar boundary well into the brown dwarf domain.

\section{Observations and Analysis}
We obtained optical spectra of the target sample using the High Resolution 
Echelle Spectrometer (HIRES; Vogt et al. 1994) on the Keck I telescope on
2002 May 19 and 20 UT. With the 1.15'' slit, the two-pixel-binned spectral 
resolution is R $\approx$ 33,000. The instrument yielded 15 spectral orders
in the 6390 -- 8700 \AA~ wavelength range, with gaps between the orders,
providing a variety of features related to youth and accretion activity. 
For comparison to our targets and to derive {\it v~sin~i}, we used M dwarf
and M giant spectroscopic standards observed with the same HIRES set up. 
The data were reduced in a standard manner using IDL routines, as described 
in Basri et al. (2000). 

We derived rotational velocities ($v~sin~i$) of the targets by 
cross-correlating with a `spun-up' template of a slowly rotating standard.
Multiple spectral orders ($\sim$ 6), selected on the basis of an absence of
strong telluric features, strong gravity-sensitive features, and stellar 
emission lines, were used in the cross-correlation analysis. 
Following White \& Basri (2002), we used a combination of giant and dwarf 
spectra for the template. 

\section{Results and Discussion}
Table 1 lists the equivalent widths of Li I and H$\alpha$ lines, full widths 
at 10\% of the peak flux level of H$\alpha$, and $v~sin~i$ values of our 
sample, along with the relevant error estimates. Of our 14 targets in Upper 
Sco, 11 show Lithium 6708 \AA~ absorption and have similar radial velocities 
(standard deviation $\sim$ 2 kms$^{-1}$).  The remaining three (USco 85, 
USco 99 and USco 121) do not show Lithium, and are likely non-members; we do 
not discuss them further. Except for GY 5, our $\rho$ Oph sources do not have
sufficient signal-to-noise in the blue to measure Li equivalent widths, 
primarily because they suffer from significant extinction. However, their
membership in the $\rho$ Oph molecular cloud has been established by
Wilking, Greene \& Meyer (1999) using near-infrared photometry and 
spectroscopy. 

Figure 1 shows H$\alpha$ line profiles of our sample. The $\rho$ Oph source 
GY 5 exhibits a broad, asymmetric H$\alpha$ 
profile while none of the Upper Sco targets do. There is evidence of a 
measurable change in the H$\alpha$ emission of USco 128 between two 
observations separated by about an hour.   

\subsection{Disk Accretion}
Accreting (``classical'') T Tauri stars exhibit strong, broad H$\alpha$
line profiles indicative of high velocities in a nearly free-falling flow
(e.g., Hartmann, Hewett, \& Calvet 1994). Weak-line T Tauri stars, on the 
other hand, harbor weak, narrow H$\alpha$ lines, presumably originating in 
their active chromospheres. The equivalent width of the H$\alpha$ line is 
often used to 
distinguish between these two types of objects. However, the threshold value 
of H$\alpha$ equivalent width depends on the spectral type (Mart\'in 1998). 
White \& Basri (2002) suggested full-width at 10\% of the peak emission as
a more accurate empirical indicator of accretion than either the H$\alpha$
equivalent width or optical veiling: 10\% widths $>$ 270 kms$^{-1}$ indicate
accretion {\it independent of the stellar spectral type}. 

The $\rho$ Oph source GY 5 has a 10\% H$\alpha$ width of 352 kms$^{-1}$
(and an equivalent width of 65 \AA~): it appears to be undergoing accretion. 
Interestingly, it is one of the sources with a mid-infrared excess detected
by the Infrared Space Observatory (ISO), consistent with the presence of 
a circumstellar disk (Comeron et al. 1998; Bontemps et al. 2001; Natta et 
al. 2002). Thus, GY 5 may be the first spectroscopically confirmed 
sub-stellar object with an accreting disk detected via infrared excess as 
well as H$\alpha$ characteristics. Surprisingly, Wilking, Greene, \& Meyer 
(1999) did not see H$\alpha$ emission ($<$ 5 \AA) in their low-resolution 
optical spectrum of GY 5 (albeit obtained ``under nonideal conditions''). 
It may be that the accretion onto GY 5 is highly variable (or highly
asymmetric so that geometric effects are important). 

According to ISO mid-infrared flux measurements, GY 141 and GY 310 also harbor
excess emission but GY 37 does not (Comeron et al. 1998; Bontemps et al. 
2001). H$\alpha$ line profiles in our optical spectra do not show clear
evidence of on-going accretion in any of these three objects. Given the 
problems of extinction, it would be interesting to obtain near-infrared 
spectra of all $\rho$ Oph sources to investigate other accretion signatures 
such as Pa $\beta$ and Br $\gamma$. Given the large ISO beam, it would also be
prudent to confirm via high angular resolution observations (e.g., ground-based
L- and N-band imaging) that the mid-infrared sources indeed coincide with the 
brown dwarfs. 

By the White \& Basri (2002) criterion, none of the Upper 
Sco objects in our sample shows evidence of on-going accretion. The case of 
USco 128, however, is intriguing. In low-resolution spectra, Ardila et al. 
(2000) reported a `constant' H$\alpha$ equivalent width of 130 \AA~ in two 
observations separated by a month. Our two HIRES spectra, separated by just 
one hour, yield equivalent widths of 16 and 25 \AA; i.e., there was a 
measurable change in the line over a short time interval (see Fig. 1). While 
it is true that low-resolution spectra of late-type objects systematically 
overestimate H$\alpha$  as a result of blending with the 6569 \AA~ TiO 
band-head, a factor of $\sim$5 difference between the Ardila et al. (2000) 
values and ours is difficult to account for in this way. However, variation 
in the H$\alpha$ width by more than a factor of 5 is not uncommon in flares. 
It may be that chromospheric activity in USco 128 is highly variable. In 
that case, the large width reported by Ardila et al. (2000) may correspond to
a period of sustained high activity and flaring. 

The mean age of stars in the Upper Sco region is $\sim$5 Myr whereas 
$\rho$ Oph sources are even younger at $\sim$1 Myr (e.g., Preibisch \&
Zinnecker 1999). Our results suggest that if most brown dwarfs are 
born with disks, their accretion rates decrease rapidly, within the first
few million years. Such a conclusion is also consistent with
measurements of disk frequency as a function of age using infrared excess. 
While a large fraction --$\sim$60\%-- of brown dwarf candidates in the 
$\sim$1-Myr-old Trapezium cluster show near-infrared excess (Muench et al. 
2001), the fraction appears to be much lower in the somewhat older $\sigma$ 
Orionis ($\sim$3-8 Myr) and TW Hydrae ($\sim$10 Myr) associations 
(Jayawardhana, Ardila \& Stelzer 2002). Thus, it appears that at least the 
inner disks of brown dwarfs deplete rather quickly, at timescales comparable 
to or smaller than those for T Tauri stars (Jayawardhana et al. 1999). 
Disk dissipation timescales in Upper Sco might be even shorter 
than in some other star-forming regions due to strong winds and ionizing 
radiation from numerous luminous stars in its midst (Preibisch \&
Zinnecker 1999 and references therein). Studies of large samples of brown 
dwarfs in several young clusters, spanning a range of ages and environments, 
will provide a more definitive answer. 

\subsection{Stellar Rotation and Activity}
Measures of stellar rotation and chromospheric activity in young brown
dwarfs, in comparison with their older counterparts in the field and 
higher-mass coeval T Tauri stars, can shed light on a variety of questions, 
from the nature of sub-stellar magnetic fields to the efficiency of disk
locking. 

A large fraction of the lowest-mass stars in Orion appear to
be fast rotators whereas Taurus objects of similar spectral type rotate
much more slowly (Clarke \& Bouvier 2000; Herbst et al. 2001; White \&
Basri 2002). According to a scenario advocated by Hartmann (2002), late-type
Orion objects have not yet had time to slow down via disk braking whereas
(somewhat older) Taurus objects have. In the standard picture, following 
disk unlocking, the low-mass stars (and presumably brown dwarfs) would spin up
again. Eventually, braking by magnetized winds is expected to slow the stars 
down again, though it appears that such braking may not be
efficient in the lowest mass stars and brown dwarfs (Mohanty \& Basri 2002a; 
Mohanty \& Basri 2002b). In the first few Myr, in any case, contraction 
timescales are expected to be much shorter than braking timescales, so 
spin-up is expected to dominate (once the disk locking ends) over this period 
(Bouvier, Forestini \& Allain 1997). 

Among the $\rho$ Oph targets, we find $ v~sin~i \approx $ 14 kms$^{-1}$ 
in GY 5, which very likely harbors a disk. It is interesting to note that
GY 141 and GY 310, which also show mid-infrared excess, are relatively slow 
rotators ($\sim$ 6 and 10 \kms, respectively) while GY 37, which does not 
show a mid-infrared excess, is a rapid rotator ($\sim$ 22 \kms). This may be 
a hint of disk-locking in action, though further observations and a larger 
sample are required for verification. 

Among the 11 Upper Sco members in our sample (all apparently non-accretors), 
six show rotational velocities $ v~sin~i >$ 15 kms$^{-1}$.  The average 
$v~sin~i$ of the Upper Sco sample is $\sim$ 25 \kms. In seven similar mass
non-accretors in Taurus, on the other hand, White \& Basri (2002) find 
an average $v~sin~i$ of $\sim$ 10 \kms, with only one object rotating 
above 15 \kms. Thus, the Upper Sco brown dwarfs appear to be rotating 
noticeably faster than their Taurus counterparts. Since Upper Sco 
is believed to be somewhat older than Taurus, we may be seeing the 
signature of spin-up due to gravitational contraction following disk 
unlocking. 

In our Upper Sco sample, the equivalent widths of H$\alpha$ arising from
chromospheric activity are in the range $\sim$ 6-18 \AA~ (except one
larger width in USco 128). Without accurate effective temperature
determinations, we cannot yet calculate the H$\alpha$ fluxes these widths 
correspond to. However, the H$\alpha$ widths are comparable to saturated 
widths in field M dwarfs of similar spectral type (Mohanty \& Basri 2002a; 
Mohanty \& Basri 2002b). Thus, if the effective temperatures of the
Upper Sco objects are similar to those of field dwarfs of the same spectral
type, then their H$\alpha$ fluxes would correspond to saturated levels
in the field. This issue will be investigated further in a subsequent paper
(Mohanty et al., in prep).

\section{Summary}
We have explored disk accretion, stellar rotation and chromospheric activity 
in a substantial sample of young brown dwarfs in the Upper Sco and $\rho$ 
Oph star-forming regions using high-resolution optical spectra from the
HIRES instrument on the Keck I telescope. We have detected evidence of
accretion in the $\rho$ Oph source GY 5 in the form of a broad, asymmetric 
H$\alpha$ emission line. Interestingly, GY 5 also has a mid-infrared 
excess detected by ISO consistent with the presence of a disk. The H$\alpha$ 
profiles of our Upper Sco targets suggest little or no on-going accretion. 
It appears that if most brown dwarfs are born with disks, their accretion 
rates decrease rapidly, on timescales comparable to or shorter than those
for T Tauri disks. On average, the Upper Sco brown dwarfs are 
rotating faster than their somewhat younger counterparts in Taurus, which 
may be a hint of spin-up due to contraction once the disk locking has ended. 
Their chromospheric activity levels, as indicated by H$\alpha$, 
are comparable to the saturated activity levels seen in field objects of
similar spectral type. Our results, in comparison with similar studies of
other young clusters spanning a range of ages and environments, will help
unveil the origin and early evolution of sub-stellar objects. 

\acknowledgements
We thank the Keck Observatory staff for their outstanding support, and 
David Ardila, Beate Stelzer and Russel White for useful conversations. 
This work was supported in part by NSF grants AST-0205130 to RJ and 
AST-0098468 to GB.

%\newpage

\clearpage
\begin{figure}
\epsscale{0.8}
\plotone{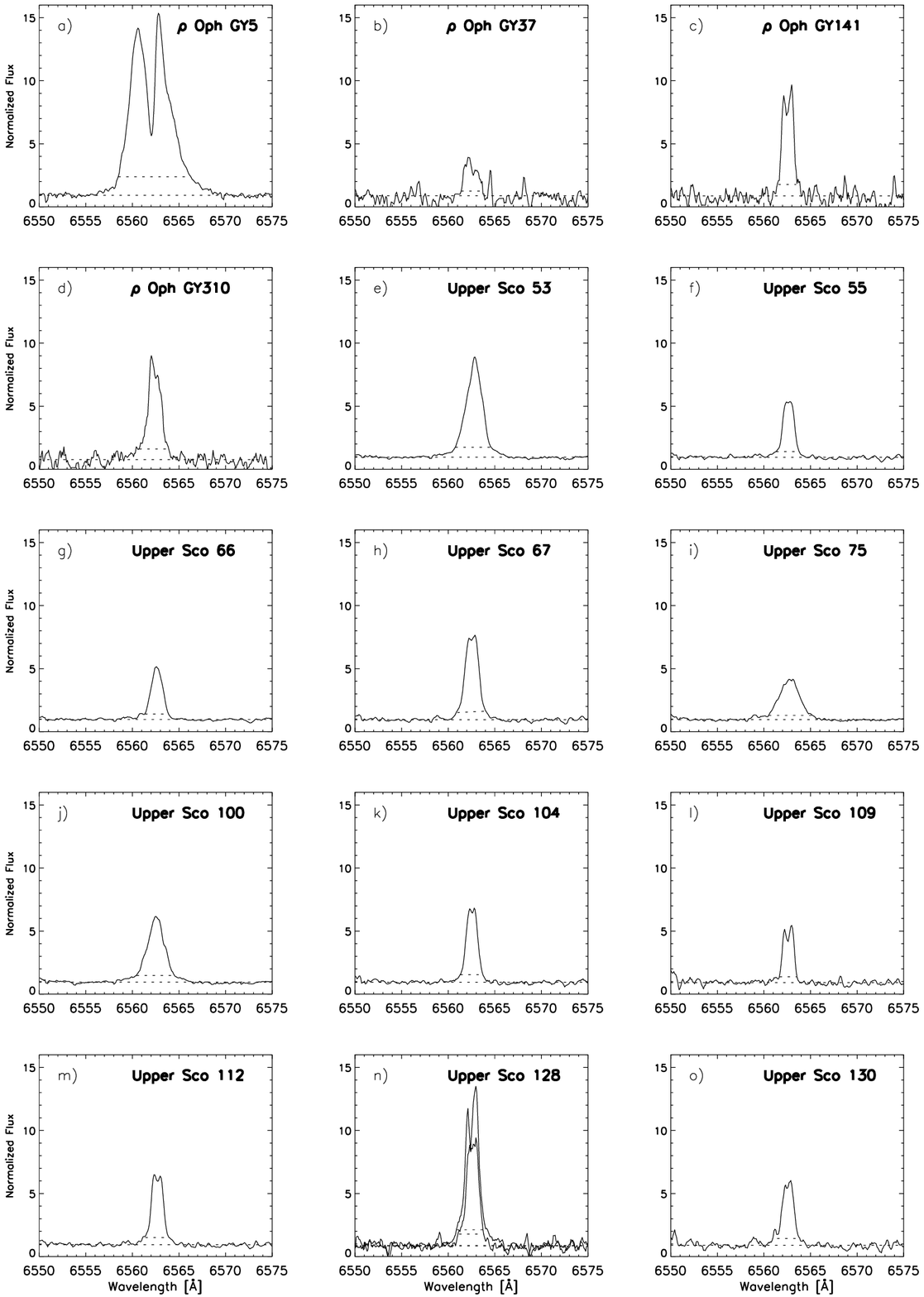}
\caption{H$\alpha$ line profiles of the target sample. Spectra shown have
been smoothed by a 3-pixel boxcar; continuum and full width at 10\% of the 
peak levels are marked by dotted lines.}
\end{figure}

\clearpage
\bigskip
\begin{center}
\begin{deluxetable}{lccccc}
\tablecaption{\label{tab1}}
\tablehead{
\colhead{Object} &
\colhead{Sp Type\tablenotemark{a}} &
\colhead{v~{\it sin}~i} &
\colhead{Li EqW\tablenotemark{b}} &
\colhead{H$\alpha$ EqW\tablenotemark{b}} &
\colhead{H$\alpha$ 10\% FW} \nl
 & & (kms$^{-1}$) & (\AA) & (\AA) & (kms$^{-1}$) \nl}

\startdata
$\rho$ Oph GY5   & M7    & 16$\pm$2   & 0.3: & 64.9$\pm$1.5 & 352 \nl
$\rho$ Oph GY37  & M6    & 22.5$\pm$5 & -    & 5.1:         & 102: \nl
$\rho$ Oph GY141 & M8.5  & 6$\pm$3    & -    & 13.4$\pm$0.2 & 87 \nl
$\rho$ Oph GY310 & M8.5  & 10$\pm$3   & -    & 17.2$\pm$0.2 & 144 \nl
USco 53          & M5    & 45$\pm$2.5 & 0.7  & 17.8$\pm$0.4 & 175 \nl
USco 55          & M5.5  & 12$\pm$3   & 0.6  & 7.3$\pm$0.3  & 114 \nl
USco 66          & M6    & 27.5$\pm$2.5 & 0.7& 6.5$\pm$0.2  & 115 \nl
USco 67          & M5.5  & 18$\pm$2   & 0.7  & 12.9$\pm$0.2 & 139 \nl
USco 75          & M6    & 62.5$\pm$5 & 0.8  & 8.9$\pm$0.2  & 212 \nl
USco 100         & M7    & 50$\pm$3   & 0.7  & 13.1$\pm$0.4 & 184 \nl
USco 104         & M5    & 16$\pm$2   & 0.6  & 9.4$\pm$0.3  & 109 \nl
USco 109         & M6    & 6$\pm$2    & 0.7  & 6.3$\pm$0.2  & 84 \nl
USco 112         & M5.5  & 5.5$\pm$2  & 0.6  & 9.5$\pm$0.3  & 111 \nl
USco 128         & M7    & $<$5\tablenotemark{c} & 0.6:\tablenotemark{c} & 15.9$\pm$0.7 & 121 \nl
                 &       &            &      & 24.8$\pm$1.7 & 147 \nl
USco 130         & M7-M8 & 14$\pm$2   & 0.6  & 8.4$\pm$0.3  & 111 \nl
\enddata
\tablenotetext{a}{Spectral types for $\rho$ Oph sources are from Wilking, Greene \& Meyer (1999); spectral types for Upper Sco targets are from Ardila et al. (2000)} 
\tablenotetext{b}{{\it Pseudo}-equivalent width; the plethora of molecular lines make it impossible to determine the true continuum} 
\tablenotetext{c}{Calculated from coadding the two spectra}
\end{deluxetable}
\end{center}

\end{document}